\documentstyle[12pt]{article}
\def\be{\begin{equation}}
\def\ee{\end{equation}}
\def\bea{\begin{eqnarray}}
\def\eea{\end{eqnarray}}

\begin{document}

\begin{center}
{\Large{\bf Some Consequences of Noncommutative Worldsheet of Superstring}}                  
										 
\vskip .5cm   
{\large Davoud  Kamani}
\vskip .1cm
 {\it Institute for Studies in Theoretical Physics and 
Mathematics (IPM)
\\  P.O.Box: 19395-5531, Tehran, Iran}\\
{\sl e-mail: kamani@theory.ipm.ac.ir}
\\
\end{center}

\begin{abstract} 

In this paper some properties of the superstring with noncommutative
worldsheet are studied. We study the noncommutativity of the spacetime, 
generalization of the Poincar\'e symmetry of the superstring, the changes
of the metric, antisymmetric tensor and dilaton.

\end{abstract} 

\newpage
\section{Introduction}
The noncommutative geometry \cite{1}
has been considered for some time in connection  
with various physics subjects. Recent motivation to study the 
noncommutative geometry mainly comes from the string theory. String theories
have been pointing towards a noncommuting scenario already in the
80's \cite{2}. Recently Yang-Mills theories on noncommutative spaces
have emerged in the context of $M$-theory compactified on a torus in
the presence of constant background three-form field \cite{3}, or as 
low-energy limit of open strings in a background $B$-field, describing
the fluctuations of the D-brane worldvolume \cite{4,5,6}.
The worldvolume of a D-brane with constant background $B$-field, is
an example of a noncommutative spacetime, in which gauge and matter fields
live \cite{3,4,5,7}. In other side, fundamental strings in the R$\otimes$R 
background $B$-field become noncommutative. This was shown in the context
of matrix theory \cite{8}.

Field theory on a noncommutative space has been proved to be 
useful in understanding various physical phenomena. Noncommutative 
field theory means that fields are thought of as  functions over 
noncommutative space. At the algebraic level, the fields become
operators acting on a Hilbert space as a representation space of
the noncommutative space \cite{9}.

If the string worldsheet lives in a noncommutative spacetime, it is
natural to expect it to inherit the noncommutativity from the spacetime.
This can be seen from the fact that the pull-back of the spacetime
noncommutativity parameter on the string worldsheet is not zero \cite{10}.
Furthermore, the string worldsheet can be a target space for another
string theory \cite{12}. Also see the Refs.\cite{13,14}. These reduce
the speculative property of the noncommutative worldsheet, and motivate
us to study it.

Previously we considered the superstring action 
as a two dimensional noncommutative field theory \cite{10}.
Up to the first order of the noncommutativity parameter, and some 
additional terms to the noncommutative superstring action,
some physical quantities are obtained. 
For example, we obtained new supersymmetric action for string, extended 
boundary state of closed superstring, new boundary conditions for 
open string which lead to the generalization of the noncommutativity 
parameter of the spacetime. 

In this paper, we do not consider the additional 
terms to the superstring action. According to this, now we study some other 
properties of the superstring with noncommutative worldsheet, to all 
orders of the noncommutativity parameter. 
Noncommutative worldsheet of string gives a noncommutative spacetime.
If some directions of the spacetime are compacted on tori, the vacuum
expectation value of the spacetime noncommutativity parameter can be non-zero.
Noncommutative worldsheet also enables us to generalize the Poincar\'e
symmetry. That is, some additional terms can be added to the ordinary
Poincar\'e transformations. Therefore, we obtain a generalized conserved
current. Furthermore, the NS$\otimes$NS fields of superstring modify by a 
common phase.
For $\theta=0$, all these results reduce to the known cases of the 
superstring theory with ordinary worldsheet, as expected.
Making use of the Refs.\cite{10,11}, one can find descriptions
of the noncommutativity parameter of the worldsheet.

This paper is organized as follows. In section 2, we study 
the action of the superstring with noncommutative worldsheet.
In section 3, the noncommutativity of the spacetime, extracted from the
noncommutativity of the string worldsheet, is studied. In section 4,
a general form of the Poincar\'e symmetry is given. 
In section 5, the effects of the noncommutativity of 
the worldsheet on the metric, Kalb-Ramond field and dilaton are extracted.
\section{Noncommutative worldsheet}
There are some models that the target space for the string theory is
two-dimensional spacetime. For example, the string worldsheet
(i.e., $\Sigma_1$) itself may emerge as the embedding space of a
{\it two-dimensional} string theory (with the worldsheet $\Sigma_2$)
\cite{12}. There may exist more primitive worldsheet theories
(i.e., $\Sigma_2$, $\Sigma_3$, ...). In this scheme the action of the
coordinates of the worldsheet $\Sigma_1$ should arise from the
two-dimensional string theory $\Sigma_2$ \cite{12}
\bea
S_1=-\frac{1}{4\pi \alpha'}
\int_{\Sigma_2} d\eta^0 d \eta^1 \sqrt{h}h^{\alpha \beta}
g_{ab}(\sigma^0 , \sigma^1)\partial_\alpha \sigma^a\partial_\beta \sigma^b,
\eea
where $\{\eta^0 , \eta^1\}$ are coordinates and $h^{\alpha \beta}$ is
metric of $\Sigma_2$. Also $g_{ab}$ is the metric of the two-dimensional
target space $\Sigma_1$.

In general, for the action (1) there are background fields \cite{13}.
These backgrounds, in the usual way, cause the coordinates
$\sigma^0$ and $\sigma^1$ to be noncommutative. This implies that in
the action of the fields $\{X^\mu(\sigma^0 , \sigma^1)\}$ the ordinary
product should change with the star product. Therefore, this also
motivates us to study the noncommutativity of the worldsheet and its
consequences.

We look at the superstring action as a two dimensional noncommutative
field theory. In other words, 
let that the superstring worldsheet be a two dimensional
noncommutative space. In this space superstring with 
worldsheet supersymmetry has the action
\bea
S_*&=&-\frac{1}{4\pi \alpha'}\int d^2 \sigma \bigg{(} 
\partial_a X^\mu * \partial^a X_{\mu}  
-i \bar{\psi}^\mu * \rho^a \partial_a \psi_{\mu} \bigg{)} \;\;,
\eea
where the spacetime and the worldsheet metrics are $\eta_{\mu \nu} =  
{\rm diag}(-1,1,...,1)$ and $\eta_{ab}={\rm diag}(-1,1)$.
We later discuss about the supersymmetry of this action.
The star product in this action
is defined between any two functions of the worldsheet 
coordinates $\sigma^a = (\sigma , \tau)$,
\bea
f(\sigma , \tau) * g(\sigma , \tau) 
= \exp \bigg{(}\frac{i}{2} \theta^{ab} \frac{\partial}{
\partial \zeta^a}\frac{\partial}{\partial \eta^b} \bigg{)} 
f(\zeta^0 , \zeta^1) g(\eta^0 , \eta^1) \bigg{|}_{\zeta^a=\eta^a=
\sigma^a}\;.
\eea
The antisymmetric tensor $\theta^{ab}$ has only one
independent component, i.e., $\theta^{ab}=\theta \epsilon^{ab}$, 
where $\epsilon^{01} = - \epsilon^{10} = 1$. Definition (3) gives 
the noncommutativity between the worldsheet coordinates as
\bea
\sigma^a * \sigma^b - \sigma^b * \sigma^a =i \theta^{ab}\;.
\eea

Usually, noncommutativity is associated with some background fields.
It seems no obvious candidate exists in two-dimension. However,
in our model the noncommutativity of the worldsheet, through the
spacetime noncommutativity, depends on the background $B$-field \cite{10}.
Furthermore, we can interpret this parameter as the UV cut-off for
the worldsheet \cite{10,11}.

The equations of motion of the worldsheet fields are
\bea
(\partial_{\tau}^2-\partial_{\sigma}^2)X^\mu 
=\partial_+ \psi_-^\mu =\partial_- \psi_+^\mu = 0\;,
\eea
where $\partial_{\pm}=\frac{1}{2}(\partial_{\tau} \pm \partial_{\sigma})$.
Therefore, $\psi^\mu_-$ and $\psi^\mu_+$ are the right moving and the 
left moving components of $\psi^\mu$. 
These equations are the same that appear for the superstring with 
ordinary worldsheet.
For the next purposes we need the solutions of these equations. For  
closed string there is
\bea
X^\mu(\sigma , \tau)= x^\mu+2\alpha'p^\mu \tau+2L^\mu\sigma+
\frac{i}{2}\sqrt{2\alpha'} \sum_{n \neq 0} \frac{1}{n} \bigg{(}
\alpha^\mu_n e^{-2in(\tau-\sigma)}+{\tilde \alpha}
^\mu_n e^{-2in(\tau+\sigma)}\bigg{)}\;,
\eea
where $L^\mu$ is zero if the direction $X^\mu$ is non-compact, and is 
$N^\mu R^\mu$ if this direction is compact on a circle with radius  
$R^\mu$. In this case the momentum of the closed string is quantized, i.e.,
$p^\mu= \frac{M^\mu}{R^\mu}$. The integers $N^\mu$ 
and $M^\mu$ are the winding and 
the momentum numbers of the closed string, respectively.

For open string we have the solution
\bea
X^\mu(\sigma , \tau)= x^\mu+2\alpha'p^\mu \tau+
i\sqrt{2\alpha'} \sum_{n \neq 0} \frac{1}{n} 
\alpha^\mu_n e^{-in\tau} \cos n\sigma \;.
\eea
Note that this solution satisfies the boundary conditions of the 
variation of the action (2). In this variation there are infinite 
number of boundary terms. These terms contain derivatives that
originate from the noncommutativity of the string worldsheet. The 
boundary condition 
\bea
(\partial_\sigma X^\mu)_{\sigma_0} = 0\;,
\eea
is sufficient to drop all the boundary terms of the above variation.
$\sigma_0 =0,\pi$ shows the boundaries of the open string worldsheet.

Consider global worldsheet supersymmetry transformations
\bea
&~&\delta X^\mu ={\bar \epsilon} \psi^\mu \;,
\nonumber\\
&~&\delta \psi^\mu = -i\rho^a \partial_a X^\mu \epsilon\;.
\eea
Invariance of the action (2) under these transformations 
leads to the worldsheet supercurrent
\bea
J_a=\frac{1}{2} \rho^b \rho_a \psi^\mu * \partial_b X_\mu \;. 
\eea
According to the equations of motion, this is a conserved current, i.e.,
$\partial^a J_a = 0 $.
\section{Spacetime noncommutativity}

Noncommutativity of the string worldsheet directly leads to the 
noncommutativity of the spacetime. Making use of the formula
\bea
e^{ip(\tau+k\sigma)}*e^{iq(\tau+l\sigma)} = e^{\frac{i}{2}\theta pq(k-l)}
e^{ip(\tau+k\sigma)}e^{iq(\tau+l\sigma)}\;,
\eea
we obtain the following noncommutativities for the spacetime
\bea
&~&[ X^\mu(\sigma , \tau)\; ,\; X^\nu(\sigma , \tau) ]_*= 
\theta (2\alpha')^{3/2} \sum_{n \neq 0}(p^\mu \alpha^\nu_n
-p^\nu \alpha^\mu_n) e^{-in\tau} \sin n\sigma 
\nonumber\\
&~&-2\alpha' \sum_{m \neq 0} \sum_{n \neq 0} \bigg{(} \frac{1}{mn}
(\alpha^\mu_n \alpha^\nu_m - \alpha^\nu_n \alpha^\mu_m )
e^{-i(m+n)\tau} \cos m (\sigma + \frac{1}{2}n \theta )
\cos n (\sigma - \frac{1}{2}m \theta )\bigg{)}\;,
\eea
from the open string point of view, and
\bea
&~&[ X^\mu(\sigma , \tau) \;,\; X^\nu(\sigma , \tau) ]_*
= 4i\theta \alpha' (p^\mu L^\nu -p^\nu L^\mu )
\nonumber\\
&~&+i\theta (2\alpha')^{3/2} \sum_{n \neq 0} \bigg{(} 
( \alpha^\mu_n p^\nu -\alpha^\nu_n p^\mu)e^{-2in(\tau-\sigma)}
-( {\tilde \alpha}^\mu_n p^\nu -{\tilde \alpha}
^\nu_n p^\mu)e^{-2in(\tau+\sigma)}\bigg{)}
\nonumber\\
&~&+2i\theta \sqrt{2\alpha'}\sum_{n \neq 0} \bigg{(} 
( \alpha^\mu_n L^\nu -\alpha^\nu_n L^\mu)e^{-2in(\tau-\sigma)}
+( {\tilde \alpha}^\mu_n L^\nu -{\tilde \alpha}
^\nu_n L^\mu)e^{-2in(\tau+\sigma)}\bigg{)}
\nonumber\\
&~& -i \alpha' \sum_{n \neq 0} \sum_{m \neq 0} \bigg{(} \frac{1}{mn}
( {\tilde \alpha}^\mu_n \alpha^\nu_m -{\tilde \alpha}
^\nu_n \alpha^\mu_m ) \sin (4mn \theta) 
e^{-2i(m+n)\tau} e^{2i(m-n)\sigma}\bigg{)}\;,
\eea
from the closed string point of view. The right hand 
sides of the relations (12) and (13) are non-zero. This is because of the
non-zero value of the parameter $\theta$. Therefore, the noncommutativity 
of the spacetime, resulted from the noncommutativity of the string
worldsheet, depends on the fact that the propagating
string in the spacetime is open or is closed.

Now consider the following expectation values of the commutators (12)
and (13)
\bea
&~&\langle 0 |[ X^\mu(\sigma , \tau) \;,\; X^\nu(\sigma , \tau) ]_* 
| 0 \rangle
= 0\; ,\;\;\;\;\;\;\;\;{\rm for\; open\; string },
\nonumber\\
&~&\langle v |[ X^\mu(\sigma , \tau) \;,\; X^\nu(\sigma , \tau) ]_* 
| v \rangle
= i \Theta^{\mu \nu} ,\;\;\; {\rm for\; closed\; string}\;,
\eea
where $|v \rangle =| 0, {\tilde 0}\;; \{M^\mu\}, \{N^\nu\} \rangle $
is a closed string state with the momentum numbers $\{M^\mu\}$ and 
the winding numbers $\{ N^\mu\}$, and
\bea
\Theta^{\mu \nu} &\equiv & 4\alpha' \theta (p^\mu L^\nu
-p^\nu L^\mu) 
\nonumber\\
&=&  4\alpha' \theta \bigg{(}
M^\mu N^\nu \frac{R^\nu}{R^\mu}
-M^\nu N^\mu \frac{R^\mu}{R^\nu} \bigg{)}\;,
\eea
therefore, if some of the directions of the spacetime are compactified 
on tori with radii $\{ R^\mu \}$, a 
closed string with noncommutative worldsheet,
momentum numbers $\{ M^\mu \}$ and winding numbers $\{ N^\mu \}$, probes
the expectation value of the noncommutativity of 
the spacetime like the relation 
(15). Open strings with noncommutative worldsheet do not probe the vacuum
expectation value of the spacetime noncommutativity.
\section{Poincar\'e symmetry}

Poincar\'e transformations $\delta X^\mu= a^\mu_{\;\;\nu} X^\nu + b^\mu$
and $\delta \psi^\mu = a^\mu_{\;\;\nu} \psi^\nu$, 
are global symmetries of the superstring theory with ordinary worldsheet, 
where $a_{\mu\nu}$
is a constant antisymmetric tensor and $b^\mu$ is a constant vector. 
Two conserved currents are associated to them. 
For the superstring theory with noncommutative worldsheet, these 
transformations can be generalized. Therefore, the effects of this
generalization also appear in the currents. The 
generalized transformations are
\bea
&~&\delta X^\mu= a^\mu_{\;\;\nu} X^\nu + b^\mu\;,
\nonumber\\
&~& \delta \psi^\mu = a^\mu_{\;\;\nu} ( \psi^\nu + \psi^\nu * \phi
-\psi^\nu  \phi )\;,
\eea
where $\phi (\sigma , \tau) $ is a scalar of the worldsheet. It will
be determined in terms of the coordinates $\sigma$ and $\tau$.

Making use of the equations of motion, the variation of the action (2)
under the transformations (16), is
\bea
\delta S_* = \frac{i}{4\pi \alpha'} a_{\mu \nu} \int 
d^2 \sigma \bigg{(} {\bar \psi}^\mu * \rho^a ( \psi ^\nu * \partial_a \phi
- \psi ^\nu \partial_a \phi) \bigg{)}\;.
\eea
For vanishing of this variation, one possibility is that $\partial_a
\phi $ be constant, i.e., independent of the coordinates $\sigma$ and $\tau$,
\bea
\partial_a \phi (\sigma , \tau) = c_a\;.
\eea
This equation has the solution 
\bea
\phi (\sigma , \tau) = c_a \sigma^a + \phi_0\;,
\eea
where $c_0$, $c_1$ and $\phi_0$ are constants.

The currents associated to the transformations (16) are
\bea
P^\mu_a = \frac{1}{2\pi \alpha'}\; \partial_a X^\mu \;,
\eea
\bea
J^{\mu \nu}_a &=& \frac{1}{4\pi \alpha'} \bigg{(} 
(X^\mu * \partial_a X^\nu - X^\nu * \partial_a X^\mu
+\partial_a X^\nu * X^\mu  - \partial_a X^\mu * X^\nu ) 
\nonumber\\
&~& +i {\bar \psi}^\mu * \rho_a ( \psi^\nu * \phi - \psi^\nu \phi)
-i {\bar \psi}^\nu * \rho_a ( \psi^\mu * \phi - \psi^\mu \phi) \bigg{)}
\nonumber\\
&~&+ i ({\bar \psi}^\mu * \rho_a \psi^\nu-
{\bar \psi}^\nu * \rho_a \psi^\mu ) \;.
\eea
The constant $\phi_0$ has no effects on the transformations (16) and on
the current (21). According to the equations of motion, these currents are
conserved, i.e., $\partial^a P^\mu_a = 0$ 
and $ \partial^a J^{\mu \nu}_a = 0$. 
If the noncommutative worldsheet changes to the ordinary worldsheet,
the transformations (16) and the current (21) reduce to the ordinary 
case, as expected.  

{\bf more generalization}

Transformations (16) can be generalized as in the following
\bea
&~&\delta X^\mu= a^\mu_{\;\;\nu} X^\nu + b^\mu\;,
\nonumber\\
&~&\delta \psi^\mu = a^\mu_{\;\;\nu} ( \psi^\nu + b_1 \psi^\nu_1
+ b_2 \psi^\nu_2+...+ b_N \psi^\nu_N)\;,
\eea
where $N$ is an integer and $\{ b_1, b_2, ..., b_N \}$ are arbitrary 
coefficients, and
\bea
&~&\psi^\mu_n = \psi^\mu_{n-1} * \phi - \psi^\mu_{n-1} \phi \;\;\;,\;\;\;
1 \leq n \leq N \;,
\nonumber\\
&~&\psi^\mu_0 = \psi^\mu \;.
\eea
Again with the choice (19) for $\phi$, the variation of the action (2) under 
the transformations (22) vanishes. Define differential operator $D$ as  
\bea
D=-\frac{1}{2}i\theta^{ab}c_a \partial_b= 
\frac{1}{2}i\theta (c_1 \partial_\tau - c_0\partial_\sigma )\;,
\eea
therefore, $\psi^\mu_n $ can be written as 
\bea
\psi^\mu_n = D^n \psi^\mu \;.
\eea
This simplifies the second transformation of (22) as in the following
\bea
\delta \psi^\mu = a^\mu_{\;\;\nu} \sum_{n=0}^N b_n D^n \psi^\nu 
\;\;,\;\;\;b_0 = 1\;.
\eea
For the special choices $b_n = \frac{1}{n!}$ and $N \rightarrow \infty $ 
this transformation becomes
\bea
\delta \psi^\mu &=& a^\mu_{\;\;\nu} \exp \bigg{(} \frac{1}{2} i \theta (c_1 
\partial_\tau - c_0 \partial_\sigma) \bigg{)} \psi^\nu (\sigma , \tau)
\nonumber\\
&=& a^\mu_{\;\;\nu} \psi^\nu \bigg{(}\sigma-\frac{i\theta}{2}c_0 \;,\;
\tau+\frac{i\theta}{2}c_1 \bigg{)}\;,
\eea
which follows by combination of the shifts on 
the worldsheet coordinates and a rotation in the spacetime. 

The currents associated to the
transformations (22), are the current (20) and 
\bea
J^{\mu \nu}_a &=& \frac{1}{4\pi \alpha'} \bigg{(} 
(X^\mu * \partial_a X^\nu - X^\nu * \partial_a X^\mu
+\partial_a X^\nu * X^\mu  - \partial_a X^\mu * X^\nu ) 
\nonumber\\
&~& +i \sum_{n=0}^N b_n ({\bar \psi}^\mu * \rho_a D^n 
\psi^\nu - {\bar \psi}^\nu * \rho_a D^n \psi^\mu) \bigg{)}\;,
\eea
which is conserved, i.e., $\partial^a J^{\mu \nu}_a =0$.
This is more general than the current (21).

Note that the parameters $\{b_n\}$, 
$c_0$ and $c_1$ in the transformation (26) and
in the current (28), remain arbitrary. For $c_0 = \pm c_1$, the operator $D$
is proportional to $\partial_{\mp}$. In this case, according to the 
equation (11), the effects of the noncommutativity of the worldsheet on the
fermionic part of the current (28), for $n \geq 1$, are collected only in the 
derivative $D$. That is, the star product appears as usual product.
Also the transformation (26) for $n \geq 1$ only has derivatives of 
the left moving or the 
right moving components of the worldsheet fermions $\{ \psi^\mu \}$.
\section{The fields $g_{\mu \nu}$, $B_{\mu \nu}$ and $\Phi$}

We are interested in to know the effects of the noncommutativity of the
string worldsheet on the metric, antisymmetric tensor and dilaton. We
discuss on these fields both in the bosonic string and in the
superstring theories.
The states of these fields can be extracted from their vertex
operators.

For the bosonic string consider the operator
\bea
\Omega^{\mu \nu} (p) = -\frac{2i}{\pi \alpha'} \int d^2 \sigma
: \partial_- X^\mu * \partial_+ X^\nu * e^{ip.X}:\;\;.
\eea
Making use of the solution (6), therefore from the state
\bea
\Omega^{\mu\nu}(0) | 0\;,{\tilde 0}\;; p=0 \rangle\;,
\eea
we read the state
\bea
e^{-4i\theta} \alpha^\mu_{-1} {\tilde \alpha}^\nu_{-1}
| 0\;,{\tilde 0}\;; p=0 \rangle\;.
\eea
According to this state we have 
\bea
&~&g^{\mu \nu}_\theta = e^{-4i\theta}g^{\mu \nu}\;,
\nonumber\\
&~&B^{\mu \nu}_\theta = e^{-4i\theta}B^{\mu \nu}\;,
\nonumber\\
&~&\Phi_\theta = e^{-4i\theta}\Phi\;.
\eea
Thus, these fields take only a phase. Modification of the dilaton
changes the string coupling constant.

For the superstring, $g_{\mu \nu}$, $B_{\mu \nu}$ and 
$\Phi$ are the NS$\otimes$NS sector
fields. The states of these fields can be extracted from the following state
\bea
-\frac{2i}{\pi} \int d^2 \sigma : \psi^\mu_- * \psi^\nu_+:
|0\;,{\tilde 0}\;; p=0 \rangle \;.
\eea
From this state we read the state
\bea
e^{-i\theta} b^\mu_{-1/2}\;{\tilde b}^\nu_{-1/2} 
|0\;,{\tilde 0}\;; p=0 \rangle\;.
\eea
According to this state, we obtain
\bea
&~&g^{\mu \nu}_\theta = e^{-i\theta}g^{\mu \nu}\;,
\nonumber\\
&~&B^{\mu \nu}_\theta = e^{-i\theta}B^{\mu \nu}\;,
\nonumber\\
&~&\Phi_\theta = e^{-i\theta}\Phi\;.
\eea
Therefore, the corrections of the metric, antisymmetric tensor and 
dilaton in the superstring theory are 
different from their corrections in the bosonic string theory. Since
$\theta$-parameter is real, the real parts of these fields can be interpreted
as physical fields.
\section{Conclusions and remarks}

The noncommutativity of the string worldsheet leads to the noncommutativity
of the spacetime. The latter depends on probing by open or 
closed string. If some of the spacetime directions 
are compactified on tori, the  
noncommutativity of the spacetime depends on the momentum numbers and the 
winding numbers of the probing closed string.

By adding some additional terms to the Poincar\'e transformations of the 
worldsheet fermions, the Poincar\'e symmetry of the superstring 
was generalized. The noncommutativity of the superstring worldsheet permits 
this generalization and consequently the generalized form of the
associated conserved current.

The NS$\otimes$NS fields of the type II superstring 
(i.e., the metric, antisymmetric tensor and dilaton)
changed only by a phase. The changes of these fields 
in the bosonic string theory, are
different from their changes in the superstring theory.

Note that the consistency of this model with two-dimensional conformal
invariance is not clear. However, for some interesting properties of it
we studied the model. If we allow the scale invariance of the worldsheet
to be broken at very short distances on the worldsheet, the UV cut-off
of the worldsheet \cite{11} can be interpreted as the parameter of the
noncommutativity \cite{10}.

\end{document}